\documentclass{SciPost}

\hypersetup{
    colorlinks,
    linkcolor={red!50!black},
    citecolor={blue!50!black},
    urlcolor={blue!80!black}
}

\newcommand{\magenta}[1]{\color{magenta} #1 \color{black}}

\usepackage[bitstream-charter]{mathdesign}
\usepackage{url}
\usepackage{color}
\usepackage{fontawesome}

\DeclareSymbolFont{usualmathcal}{OMS}{cmsy}{m}{n}
\DeclareSymbolFontAlphabet{\mathcal}{usualmathcal}

\fancypagestyle{SPstyle}{
\fancyhf{}
\lhead{\colorbox{scipostblue}{\bf \color{white} ~SciPost Physics Community Reports }}
\rhead{{\bf \color{scipostdeepblue} ~Submission }}

\fancyfoot[C]{\textbf{\thepage}}
}

\begin{document}\vspace*{-1cm}
\begin{flushleft} 
\magenta{LHCHWG-2026-004}\\
\magenta{IFT-UAM/CSIC-26-65}
\end{flushleft}

\pagestyle{SPstyle}

\begin{center}{\Large \textbf{\color{scipostdeepblue}{
%%%%%%%%%% TODO: Write your article's title here
NLO EW and QCD dimension-6 SMEFT results for Higgs and gauge boson decays in POPxf format\\
}}}\end{center}

\begin{center}\textbf{
% Use (full) first name (+ middle name initials) + surname format.
% Separate subsequent authors by a comma, omit comma and use "and" for the last author.
% Mark the corresponding author(s) with a superscript symbol in this order
% \star, \dagger, \ddagger, \circ, \S, \P, \parallel, ...
% Aah B. Cee\textsuperscript{1},
% Dee E. Faa\textsuperscript{2} and
% Gee K. See\textsuperscript{3$\dagger$}
Luigi Bellafronte\textsuperscript{1$\star$},
Sally Dawson\textsuperscript{2$\dagger$},
Clara Del Pio\textsuperscript{2$\ddagger$},\\
Matthew Forslund\textsuperscript{3$\circ$} and
Pier Paolo Giardino\textsuperscript{4$\S$}
}\end{center}

\begin{center}
{\bf 1} Physics Department, Florida State University, Tallahassee, FL 32306-4350, USA
\\
{\bf 2} High Energy Theory Group, Physics Department, 
    Brookhaven National Laboratory, Upton, NY 11973, USA
\\
{\bf 3} Princeton Center for Theoretical Science, Princeton University, Princeton, NJ, 08544 USA
\\
{\bf 4} Departamento de Física Teórica and Instituto de Física Teórica UAM/CSIC, 
    Universidad Autónoma de Madrid, Cantoblanco, 28049, Madrid, Spain
%%%%%%%%%% END TODO: AFFILIATIONS
% Provide email address of corresponding author(s)
\\[\baselineskip]
$\star$ \href{mailto:lbellafronte@fsu.edu}{\small lbellafronte@fsu.edu}\,,\quad
$\dagger$ \href{mailto:dawson@bnl.gov}{\small dawson@bnl.gov}\,,\quad
$\ddagger$ \href{mailto:cdelpio@bnl.gov}{\small cdelpio@bnl.gov}\,,\\
$\circ$ \href{mailto:mforslund@princeton.edu}{\small mforslund@princeton.edu}\,,\quad
$\S$ \href{mailto:pier.giardino@uam.es}{\small pier.giardino@uam.es}
\end{center}

\section*{\color{scipostdeepblue}{Abstract}}
\textbf{\boldmath{%
We present next-to-leading-order (NLO) QCD and electroweak (EW) results  using the dimension-6 SMEFT for all 2- and 4- body Higgs decays, for $Z$ and $W$ decays along with the corresponding EW precision observables, and for the Higgstrahlung process $e^+e^-\rightarrow ZH$ at $\sqrt{s}=240$, $365$ and $500$  GeV.  The results are presented in the POPxf format for ease of use in experimental and phenomenological studies.  Of particular utility is the total Higgs width, including all dimension-6 contributions at NLO. In addition, we present the differential distributions $d\Gamma/dm_{Z*}$  for $H\rightarrow l^+l^- Z^*, Z^*\rightarrow l^+l^-$ at NLO in the SMEFT.
}}

\vspace{\baselineskip}

%%%%%%%%%% BLOCK: Copyright information
% This block will be filled during the proof stage, and finilized just before publication.
% It exists here only as a placeholder, and should not be modified by authors.
\noindent\textcolor{white!90!black}{%
\fbox{\parbox{0.975\linewidth}{%
\textcolor{white!40!black}{\begin{tabular}{lr}%
  \begin{minipage}{0.6\textwidth}%
    {\small Copyright attribution to authors. \newline
    This work is a submission to SciPost Physics. \newline
    License information to appear upon publication. \newline
    Publication information to appear upon publication.}
  \end{minipage} & \begin{minipage}{0.4\textwidth}
    {\small Received Date \newline Accepted Date \newline Published Date}%
  \end{minipage}
\end{tabular}}
}}
}
%%%%%%%%%% BLOCK: Copyright information

% For convenience during refereeing we turn on line numbers:
%\linenumbers

\vspace{10pt}
\noindent\rule{\textwidth}{1pt}
\tableofcontents
\noindent\rule{\textwidth}{1pt}
\vspace{10pt}

\section{Introduction}
\label{sec:intro}
One of the major goals of the high-luminosity LHC is to precisely limit  or to observe deviations from the Standard Model (SM) predictions.  The Standard Model Effective Field Theory (SMEFT)~\cite{Brivio:2017vri} is a useful tool for searching for high scale new physics effects not present in the SM and postulates that there are no unobserved light particles and that the gauge symmetry remains the $SU(3)\times SU(2)\times U(1)$ symmetry of the SM.  The Lagrangian is then expanded around the SM,
\begin{align}
{\mathcal{L}}={\mathcal{L}}_{SM}+\sum_{i,d}\frac{{\hat C}_i^{(d)}}{ \Lambda^{d-4}}
{\mathcal{O}}_i^{(d)}\, ,
\end{align}
where ${\mathcal{O}}_i^{(d)}$ are the complete set of $SU(3)\times SU(2)\times U(1)$ invariant dimension-$d$ operators constructed out of SM  fields and $\hat{ C}_i^{(d)}$ are the corresponding Wilson coefficients, which we take to be in the Warsaw basis~\cite{Grzadkowski:2010es,Dedes:2017zog}.  We restrict ourselves to dimension-$6$ operators. The expansion parameter, $\Lambda$, is
assumed to be TeV scale or higher.

Predictions for  physical observables, $O_\alpha$, are expanded around the SM predictions as an inverse power series in $\Lambda$  and in the loop expansion, ${1\over 16\pi^2}$,
\begin{align}
O_\alpha=O_{\alpha,SM}+\sum_{i,j,d}\beta_{\alpha,i,j}{{\hat C}_i^{(d)}\over (\Lambda^{d-4})(16 \pi^2)^j}
\end{align}
Tree level observables ($j=0$) accurate to dimension-6 ($d=6$) can be computed in a straightforward manner using { \sc MadGraph}~\cite{Alwall:2014hca} in conjunction with SMEFTsim~\cite{Brivio:2020onw} or SMEFT@NLO~\cite{Degrande:2020evl}.  
We retain terms only to ${\cal{O}}({1\over\Lambda^2})$ since a consistent tree level calculation to ${\cal{O}}({1\over\Lambda^4})$ would require including the contributions of dimension-8 operators.

The anticipated precision from the HL-LHC, along with projections for FCC-ee, require that the SMEFT predictions be computed beyond the leading order in the loop expansion.  Next-to-leading-order (NLO) QCD corrections including the contributions of dimension-6 operators can be found from the automated tools listed above.  NLO electroweak (EW) corrections, however, present a different challenge from QCD corrections, as they typically introduce a dependence on dozens of Wilson  coefficients that do not contribute at tree level and are calculated on a case-by-case basis, with the results  scattered through the literature.
The NLO calculations we describe in this note are accurate to ${\cal{O}}({1\over 16\pi^2\Lambda^2})$, as reaching 
the accuracy ${\cal{O}}({1\over 16\pi^2\Lambda^4})$ would require the inclusion of one-loop diagrams with two insertions of dimension-6 operators, along with dimension-8 operators, which is beyond the scope of the results presented here.

The numerical results of the various existing dimension-6 NLO QCD/EW calculations can be conveniently summarized using the POPxf approach~\cite{Brivio:2025mww}.  
The POPxf exchange format is a recently proposed data structure designed to standardize arbitrary polynomial dependences of observables on model parameters to simplify sharing and reproducing theoretical predictions in effective field theory (EFT) analyses. 
The format supports observables expressed as arbitrary functions of polynomials of any degree, and also allows the inclusion of uncertainties and correlations.
Additional metadata entries include all relevant information for reproducibility, including renormalization scales, parameter inputs, choice of SMEFT basis, etc.

Our usage of this  formalism includes the numerical results for $\beta_{\alpha,i,0}$ and $\beta_{\alpha,i,1}$ in JSON files which include sufficient information such that the results can be replicated by the user. 
Our observables for Higgs decays consist of inclusive decay widths, branching ratios, and some differential distributions for $H\to4\ell$ that will be discussed below.
While the POPxf format allows arbitrary degree polynomials, to maintain a consistent power counting all our results are expanded only up to linear $\mathcal{O}(\frac{1}{\Lambda^2})$ order.
In this note, we compile and present NLO results for Higgs decays~\cite{Bellafronte:2026mhp, Bellafronte:2025jbk}, W and Z decays along with the corresponding electroweak precision observables~\cite{Bellafronte:2023amz,Dawson:2019clf,Biekotter:2025nln}, and the Higgstrahlung process, $e^+e^-\rightarrow ZH$~\cite{Asteriadis:2024xuk,Asteriadis:2024xts}.
Results are presented with $\Lambda = 1$ GeV such that all coefficients are in units of GeV$^{-2}$, following POPxf conventions for the "Warsaw" WCxf basis.
The POPxf files are located at a \href{https://gitlab.com/mforslund/smeft-nlo-popxf}{GitLab~\faGitlab} repository.

\section{Processes}

As described in~\cite{Bellafronte:2025jbk,Bellafronte:2026mhp}, all of the two- and four-body Higgs decays and branching ratios are now known to NLO QCD/EW in the dimension-$6 $ SMEFT in the narrow width approximation:
\begin{align}
    &H\rightarrow \gamma\gamma , \quad H\rightarrow \gamma Z , \quad 
    H\rightarrow gg\nonumber\\
    &H\to f\bar{f} \, , \quad f \in [b,\tau,\mu,c,s]\\ 
    &H\rightarrow (f_{g_1}\overline{f}_{g_1})(f_{g_2}\overline{f}_{g_2})\, , \quad f \in [\ell,\nu_\ell,u,d]
\end{align}
where $g_k=1,2,3$ is a generation index.
These processes are implemented in the public code {\sc NEWiSH}~\cite{GITLAB:newish}.
Our default set of input parameters for Higgs decays is $\{ M_W,\, M_Z,\, G_F\}$. The couplings and gauge boson masses are renormalized in the on-shell scheme. We  present results using both the on-shell and the ${\overline{\mathrm{MS}}}$ renormalization scheme for fermion masses.  
 All results have an arbitrary flavor scheme, which can easily be restricted to a desired flavor assumption. 
 The results for the total Higgs decay widths, along with numerical values of the input parameters can be found
 in POPxf notation at the GitLab repository~\cite{GITLAB:nlo_smeft_popxf}.  
 In addition, for the 4-body Higgs decays to leptons, we also include distributions for $H\rightarrow 4\ell$ in the GitLab repository.

The NLO QCD/EW dimension-6 results for $Z$ and $W$ decays have also been implemented in POPxf formalism in both the $\{ M_W,\, M_Z,\, G_F\}$ and $\{ \alpha(0),\,M_Z,\, G_F\}$ input schemes, along with the EW precision observables (EWPOs),\footnote{In the $\{M_W,M_Z, G_F\}$ input scheme, $M_W$ is replaced by $\alpha$ in Eq. \eqref{eq:ewpoobs}.  Definitions of the observables can be found in \cite{Dawson:2019clf}.}
\begin{align}
\alpha,\Gamma_W(\textrm{total}),\Gamma_Z(\textrm{total}),R_e,R_\mu,R_\tau,R_c,\sigma(\textrm{had}),A_\mu,A_\tau,A_s,A_c,A_b,A_{FB}^e,A_{FB}^\tau,A_{FB}^s,A_{FB}^c, A_{FB^b}\, .
\label{eq:ewpoobs}
\end{align}These results are particularly useful for deriving sensitivities at a future Tera-Z facility~\cite{Bellafronte:2023amz,Dawson:2019clf,Biekotter:2025nln}.

For completeness, we also include results for $e^+e^-\rightarrow ZH$~\cite{Asteriadis:2024xts,Asteriadis:2024xuk}.
We include both polarized and unpolarized results at $\sqrt{s}=240$, $365$, and $500$ GeV with renormalization scales $\mu=\sqrt{s}$ and $\mu=1$ TeV.

\section{Results}

\subsection{Higgs Decays}\label{sec:HiggsDecays}
% We parameterize the Higgs decay widths and branching ratios as, 
% \begin{align}
% {\Gamma(H\rightarrow X)^{(0,1)}\over \Gamma(H\rightarrow X)_{\mathrm{SM}}}=1+\sum_i \delta_i^{(0,1)}{C_i}\\
% {BR(H\rightarrow X)^{(0,1)}\over BR(H\rightarrow X)_{\mathrm{SM}}}=1+\sum_i \Delta_i^{(0,1)}{C_i}
% \label{eq:expa}
% \end{align}
% where $\Gamma_{\mathrm{SM}}$  and $BR(H\rightarrow X)_{\mathrm{SM}}$ are the world's best theory calculations taken from~\cite{LHCHiggsCrossSectionWorkingGroup:2016ypw}, the superscripts $(0,1)$ denote a LO or NLO result, respectively, and we have absorbed the factor of $\frac{1}{\Lambda^2}$ into the definition of $C_i\equiv \frac {\hat {C}_i^{(6)}}{\Lambda^2}$.
We parameterize the Higgs decay widths as\footnote{Note that this choice is different from the results presented in~\cite{Bellafronte:2025jbk,Bellafronte:2026mhp} where the widths were not rescaled.}
\begin{align}
\tilde{\Gamma}(H\to X)\equiv\Gamma(H\to X)_\text{SM} \times{\Gamma(H\rightarrow X)^{(0,1)}\over \Gamma(H\rightarrow X)^{(0,1)}_{\mathrm{SM}}}= \Gamma(H\to X)_\text{SM} \times(1+\sum_i \delta_i^{(0,1)}{C_i})
\end{align}
where the superscripts $(0,1)$ denote a LO or NLO result, respectively, $\Gamma_{\mathrm{SM}}^{(0,1)}$ is our SM prediction, and $\Gamma_{\mathrm{SM}}$ is the world's best prediction taken from~\cite{LHCHiggsCrossSectionWorkingGroup:2016ypw}. We have absorbed the factor of $\frac{1}{\Lambda^2}$ into the definition of $C_i\equiv \frac {\hat {C}_i^{(6)}}{\Lambda^2}$ with $\Lambda = 1$ GeV such that $C_i$ are in units of GeV$^{-2}$.
We define the total width to be the sum of the rescaled widths:
\begin{equation}
    \Gamma_\text{Total} = \sum_X \tilde{\Gamma}(H\to X)
\end{equation}
With this definition, we define the branching ratio to be the ratio of rescaled widths parameterized as
\begin{equation}\label{eq:expa}
    BR(H\rightarrow X)^{(0,1)} = \frac{\tilde{\Gamma}(H\to X)}{{\Gamma_\text{Total}}}=BR(H\to X)_\text{SM} \times(1+\sum_i \Delta_i^{(0,1)}{C_i})
\end{equation}
where we have expanded the ratio to $\mathcal{O}(\frac{1}{\Lambda^2})$. 
This rescaling procedure helps mitigate the impact of quark mass scheme choice and higher order SM corrections on our SMEFT predictions.
We take a renormalization scale $\mu=m_H$, and take for the remaining inputs
\begin{align}
\begin{split}
    M_W  &= 80.352~\textrm{GeV} \\
    M_Z  &= 91.1535~\textrm{GeV} \\
    G_\mu &= 1.16638\times 10^{-5}~\textrm{GeV}^{-2} \\
    m_H &= 125.1~\textrm{GeV}\end{split}
    \begin{split}
    m_t &= 172.76~\textrm{GeV}\\
    \alpha_S(m_) &= 0.1188 \\
    \Delta \alpha^{(5)}_\textrm{had} &= 0.02768 \, ,
\end{split}
\end{align}
where the $W$ and $Z$ masses correspond to experimental values $M_W^\mathrm{exp}=80.379$ 
GeV and $M_Z^\mathrm{exp}=91.1876$ GeV, rescaled using the pole definitions
\begin{equation}
    M_V=\frac{M_V^\mathrm{exp}}{\sqrt{1+\left(\Gamma_V^\mathrm{exp}/{M_V^\mathrm{exp}}\right)^2}} \, ,
\end{equation}
where $\Gamma^\mathrm{exp}_W=2.085$ GeV and $\Gamma^\mathrm{exp}_Z = 2.495$ GeV.
For $H\to f\bar{f}$ decays where light fermion masses are included, we take as inputs the on-shell (OS) and ${\overline{MS}}$ quark masses,
\begin{align}
\begin{split}
    m_\tau &= 1.777~\textrm{GeV}\\
    m_b^\mathrm{OS} &= 4.92~\textrm{GeV}\\
    m_c^\mathrm{OS} &= 1.51~\textrm{GeV}\\
    m_s^\mathrm{OS} &= 0.1~\textrm{GeV}\, ,
\end{split}
\begin{split}
    m_\mu &= 0.1057~\textrm{GeV}\\
    m_b^{\overline{\mathrm{MS}}} (m_b) &= 4.183~\textrm{GeV}\\
    m_c^{\overline{\mathrm{MS}}} (m_c) &= 1.273~\textrm{GeV}\\
    m_s^{\overline{\mathrm{MS}}} (2  \ \mathrm{GeV})&= 0.0935~\textrm{GeV}\, .
\end{split}
\end{align}
where the ${\overline{\mathrm{MS}}}$ quark masses are run to $\mu=m_H$ using {\sc DsixTools}~\cite{Celis:2017hod,Fuentes-Martin:2020zaz}.
For $H\to 4f$, we compute the full four-body final state at LO, while in the NLO correction we use the narrow width approximation, $H \to f_{g_1}\bar{f}_{g_1}V, \, V \to f_{g_2}\bar{f}_{g_2}$.
Inclusive numerical values for $\delta_i$ and $\Delta_i$ are in the POPxf files.  

\begin{figure*}[t]
\centering
    \includegraphics[width=0.8\textwidth]{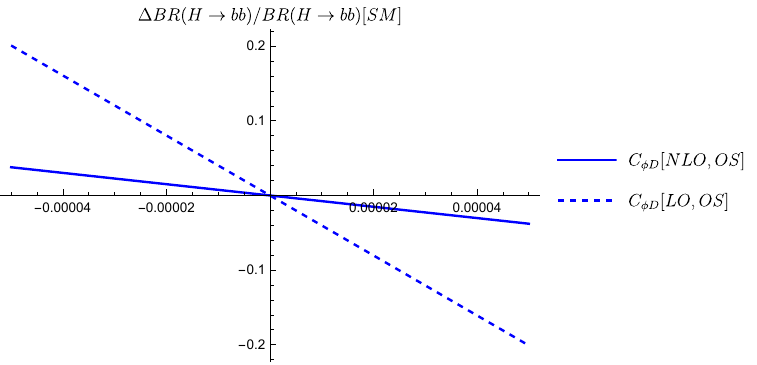}
	\caption{ SMEFT contributions to the branching ratio of Higgs decay to $b {\overline{b}}$ normalized to the best SM value.  The input scheme is $\{M_W, M_Z, G_F\}$, with a comparison of on-shell and ${\overline{\mathrm{MS}}}$ renormalization of the masses. The difference between the OS and ${\overline{MS}}$ schemes is negligible.} 
    \label{fg:hbrb}
\end{figure*}

\begin{figure*}[t]
	\centering
\includegraphics[width=0.8\textwidth]{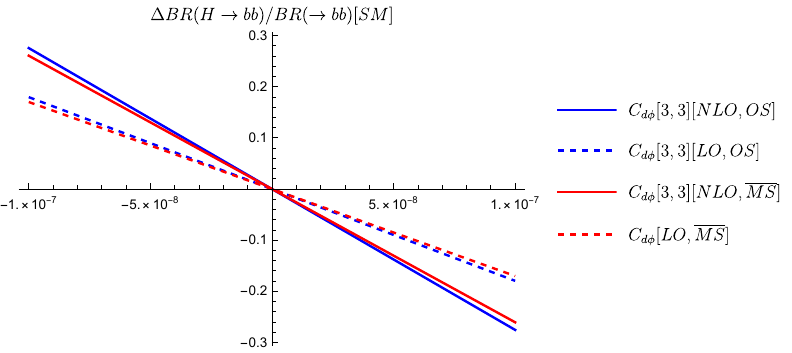}
	\caption{ SMEFT contributions to the branching ratio of Higgs decay to $b {\overline{b}}$ normalized to the best SM value. The input scheme is $\{M_W, M_Z, G_F\}$, with a comparison of on-shell and ${\overline{\mathrm{MS}}}$ renormalization of the masses.  }
    \label{fg:hbrb2}
\end{figure*}

\begin{figure*}[h]
	            \centering
         \includegraphics[width=0.8\linewidth]{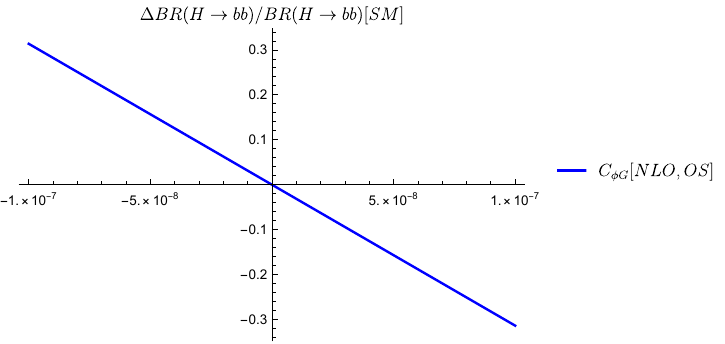}
	            \caption{SMEFT contribution to the branching ratio of Higgs decay to $b {\overline{b}}$ normalized to the best SM value.  The input scheme is $\{M_W, M_Z, G_F\}$. $C_{\phi G}$ enters only through its contribution to the total width and the LO and NLO curves are indistinguishable. The difference between the OS and ${\overline{MS}}$ schemes is also negligible. }
	        \label{fg:cphig}
	        \end{figure*}

\begin{figure*}[t]
	\centering  
\includegraphics[width=0.8\textwidth]{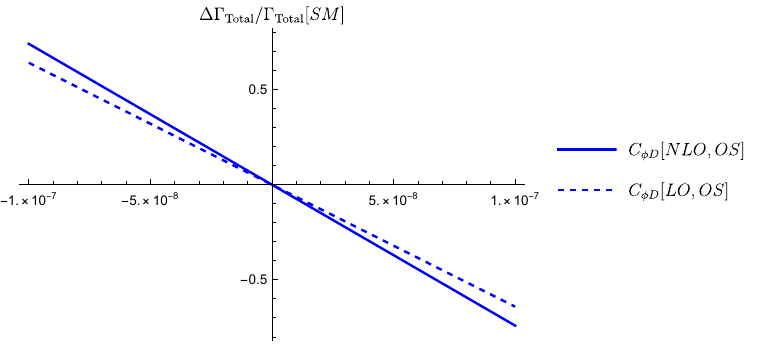}
	\caption{SMEFT contributions to the total Higgs width normalized to the best SM value.   The difference between the OS and ${\overline{MS}}$ schemes is negligible.
   } 
	\label{fg:tot}
\end{figure*}

\begin{figure*}[t]
	\centering  
\includegraphics[width=0.8\textwidth]{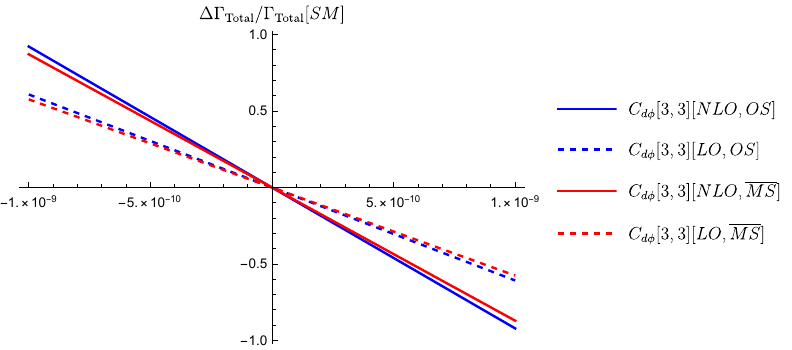}
	\caption{SMEFT contributions to the total Higgs width normalized to the best SM value.   There are small differences between the OS and ${\overline{\mathrm{MS}}}$  schemes. 
   } 
	\label{fg:tot2}
\end{figure*}

\begin{figure*}[t]
	\centering  
\includegraphics[width=0.8\textwidth]{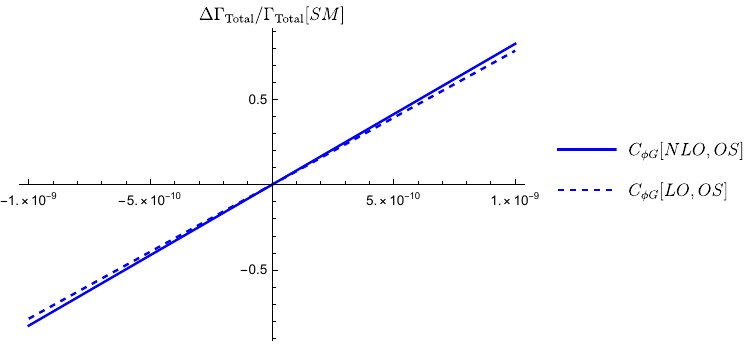}
	\caption{SMEFT contributions to the total Higgs width normalized to the best SM value. The difference between the  OS and ${\overline{MS}}$ schemes is negligible. } 
	\label{fg:totcphig}
\end{figure*}

For  $H\to q\bar{q}$, the choice of the on-shell vs $\overline{\mathrm{MS}}$ renormalization scheme  for quark masses can have a sizable impact on the numerical results, depending on the operator.
We show in Figs. \ref{fg:hbrb}  and \ref{fg:hbrb2} an example of this scheme dependence for $C_{\phi D}$ and $C_{d \phi}[3,3]$.  The dependence on $C_{\phi D}$ has sizable cancellations at LO, making the difference between LO and NLO larger than for most other coefficients. For  many operators, the scheme dependence on the quark masses is negligible, as  seen in  Figs. \ref{fg:hbrb}-  \ref{fg:cphig}.
Since $H\to b\bar{b}$ is the largest branching ratio, the scheme dependence  translates into scheme dependence of all branching ratios when expanding the denominator, as we show in Figs.  \ref{fg:tot} - \ref{fg:totcphig}.
Both mass renormalization schemes are included as separate POPxf files.

\begin{figure*}[h]
	\centering  
\includegraphics[width=0.8\textwidth]{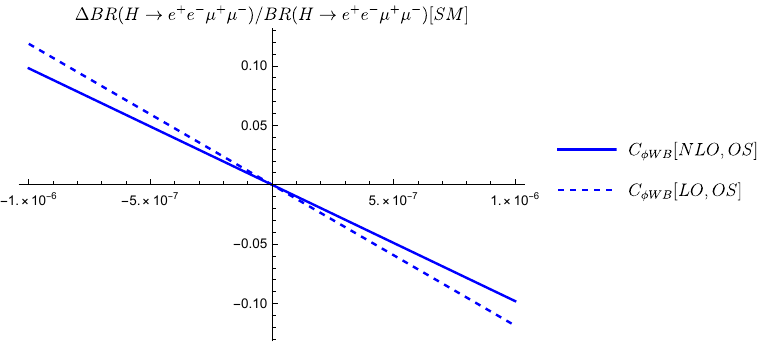}
	\caption{SMEFT contribution to the branching ratio $H\rightarrow e^+e^- \mu^+\mu^-$. The curves with OS and ${\overline{\mathrm{MS}}}$ renormalization are indistinguishable. } 
	\label{fg:4ltot}
\end{figure*}

\begin{figure*}[h]
	\centering  
\includegraphics[width=0.8\textwidth]{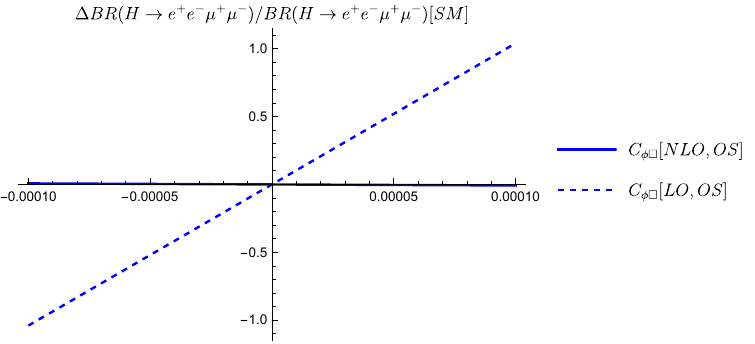}
	\caption{SMEFT contribution to the branching ratio $H\rightarrow e^+e^- \mu^+\mu^-$.  The difference between the OS and ${\overline{MS}}$ schemes is negligible. At LO, there is a large cancellation at ${\cal{O}}({1\over \Lambda^2})$ between $\Gamma(H\rightarrow b {\overline {b}})$ and $\Gamma_{\textrm{Total}}$ for $C_{\phi\square}$ when computing the branching ratio.  } 
	\label{fg:4ltot2}
\end{figure*}

An example of the $H\rightarrow e^+e^- \mu^+\mu^-$  branching ratio is shown in Fig. \ref{fg:4ltot} for the $C_{\phi WB}$ contribution which appears at LO  and we see a modest shift at NLO.  The $C_{\phi\square}$ contribution \ref{fg:4ltot2} illustrates the dependence on the mass renormalization scheme.

\begin{figure*}[h]
	\centering  
\includegraphics[width=0.8\textwidth]{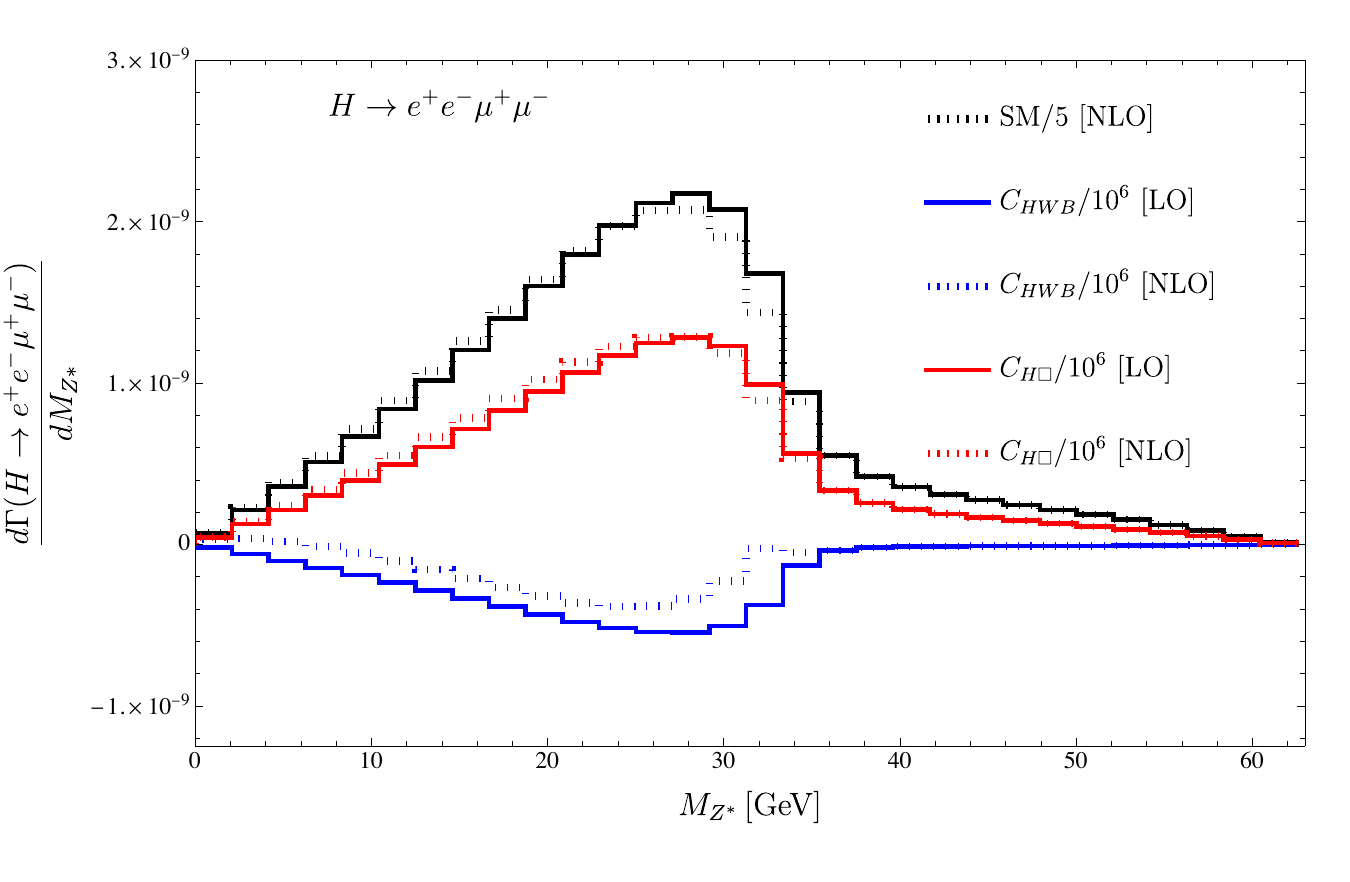}
	\caption{SMEFT contributions to the $d\Gamma /dM_{Z^*}$ differential distributions for $H\rightarrow e^+e^- \mu^+\mu^-$.  }
	\label{fg:dists}
\end{figure*}

\begin{figure*}[h]
	\centering  
\includegraphics[width=0.8\textwidth]{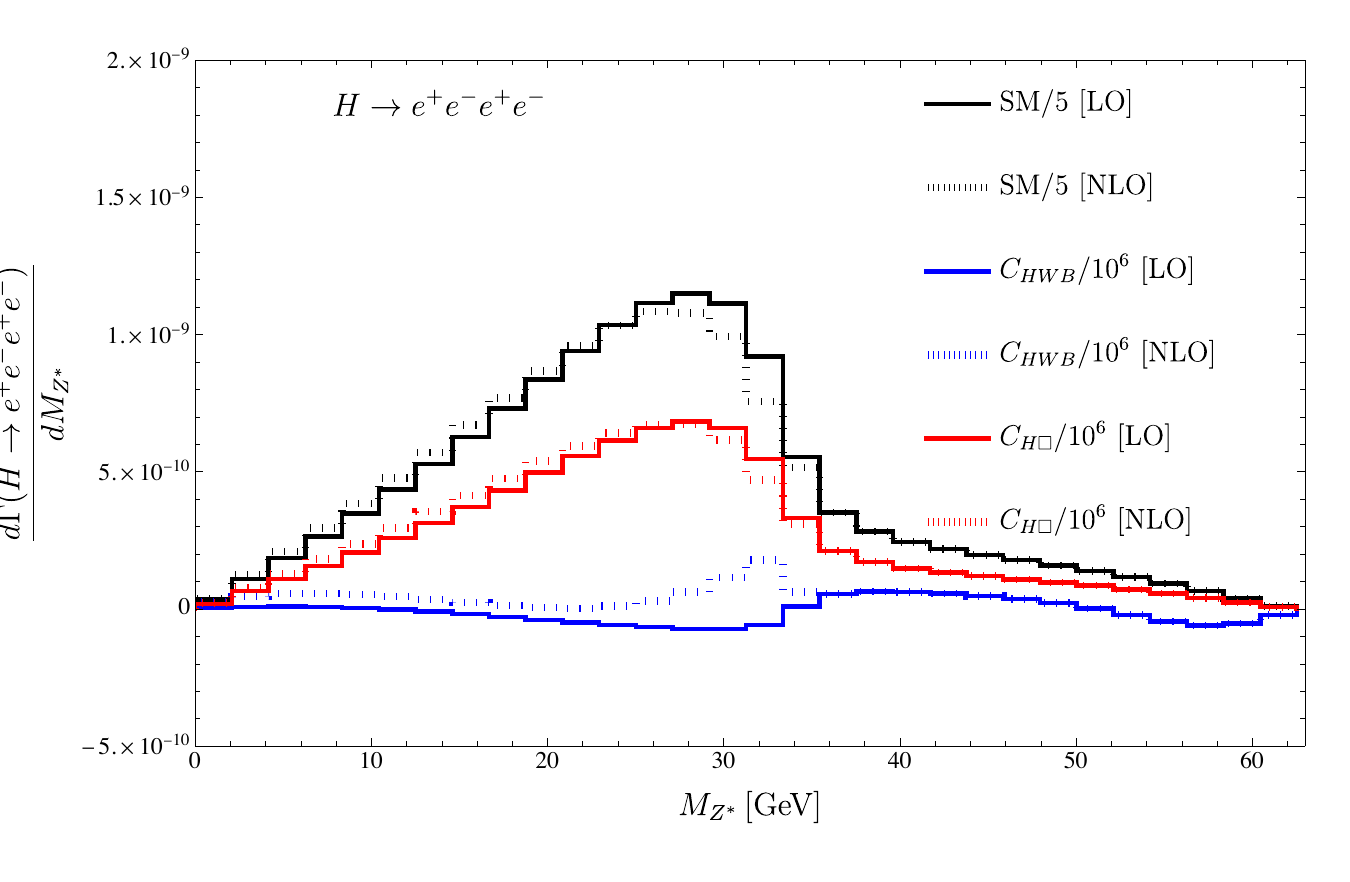}
	\caption{SMEFT contributions to the $d\Gamma /dM_{Z^*}$ differential distributions for  $H \rightarrow e^+e^- e^+e^-$.  }
	\label{fg:dists2}
\end{figure*}

For $H \to 4\ell$, we include files with a cut of $M_{Z^*}>12$ GeV motivated by experimental analyses~\cite{ATLAS:2023tnc,CMS:2025wnr,Dawson:2024pft}.
Here we define $M^{\ell\ell}_Z$ to be the same-flavour opposite sign lepton pair with invariant mass $M_{\ell\ell}$ closest to $M_Z$, and $M_{Z^*}$ the opposite lepton pair.
We also include results for $d\Gamma/dM_{Z^*}$ differential distributions for $H\to e^+e^-e^+e^-$, $H\to e^+e^-\mu^+\mu^-$, and $H\to \mu^+\mu^-\mu^+\mu^-$.
We illustrate these results in Figs.~\ref{fg:dists} and \ref{fg:dists2} for a few select coefficients.\footnote{Note that in the $M_{Z^*}>12$ GeV inclusive results as well as the $d\Gamma/dM_{Z^*}$ differential distributions we use our own NLO SM calculation rather than a world's best calculation.}

For further details of the NLO calculation for all of the $H\to X$ processes including renormalization, IR subtraction, and a discussion of the validity of the narrow width approximation, we refer to the original publication~\cite{Bellafronte:2026mhp} as well as the README at the GitLab repository~\cite{GITLAB:nlo_smeft_popxf}.

\subsection{Electroweak Observables}\label{sec:ZPole}

We have included NLO QCD and EW results  for EWPOs, which provide significant constraints on dimension-6 global SMEFT fits, in the POPxf format. All 2-body decays of $Z$ and $W$ bosons at NLO  are known to linear order in the dimension-6 SMEFT, with no assumptions on the flavor structure, allowing for  the construction of the EWPOs.

\begin{figure*}
\centering
	 \includegraphics[width=0.8\textwidth]{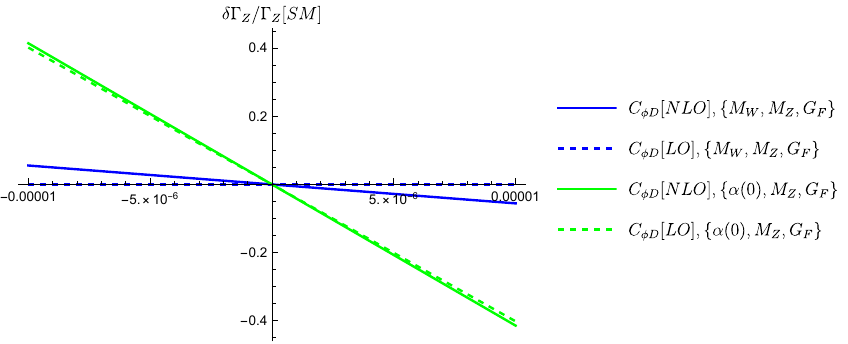}
	\caption{Results for $\Gamma_Z$ proportional to $C_{\phi D}$ at LO and NLO in different input schemes normalized to the most accurate theoretical prediction. }
    \label{fg:gamz}
    \end{figure*}

\begin{figure*}
	\centering      \includegraphics[width=0.8\textwidth]{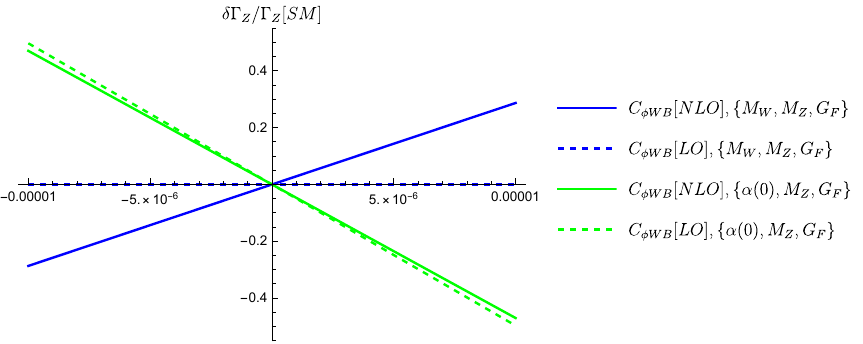}
	\caption{Results for $\Gamma_Z$ proportional to  $C_{\phi WB}$  at LO and NLO in different input schemes normalized to the most accurate theoretical prediction.
    } 
    \label{fg:gamz2}
    \end{figure*}

\begin{figure*}
	\centering  
      \includegraphics[width=0.8\textwidth]{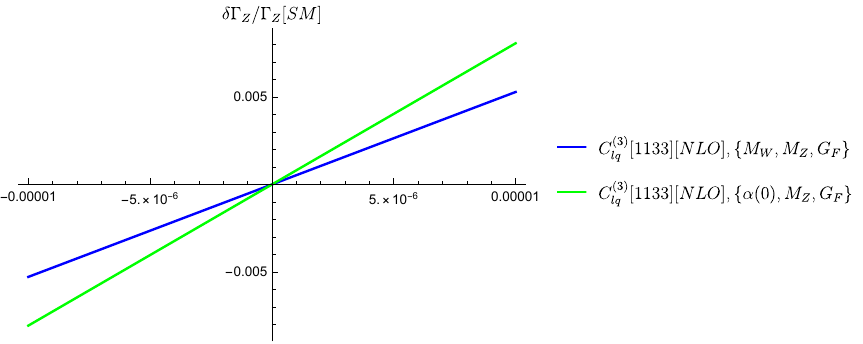}
	\caption{Results for $\Gamma_Z$ proportional to $C_{lq}^{(3)}$ at LO and NLO in different input schemes normalized to the most accurate theoretical prediction.
    } 
    \label{fg:gamz3}
    \end{figure*}

\begin{figure*}
	\centering  
      \includegraphics[width=0.8\textwidth]{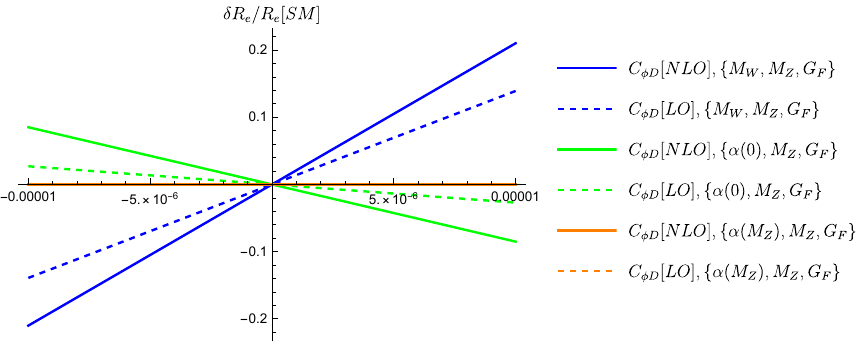}
	\caption{Results for $R_e$ proportional to $C_{\phi D}$ at LO and NLO in different input schemes normalized to the most accurate theoretical prediction. 
    } 
    \label{fg:Re}
    \end{figure*}

\begin{figure*}
	\centering  
      \includegraphics[width=0.8\textwidth]{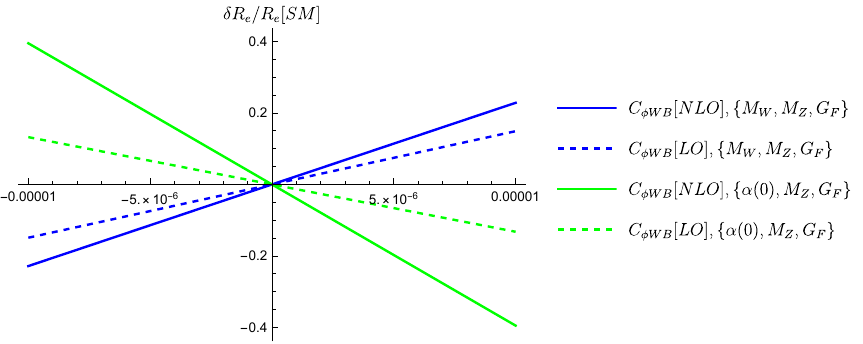}
	\caption{Results for $R_e$ proportional to  $C_{\phi WB}$ at LO and NLO in different input schemes normalized to the most accurate theoretical prediction.  } 
    \label{fg:Re2}
    \end{figure*}

\begin{figure*}
	\centering  

      \includegraphics[width=0.8\textwidth]{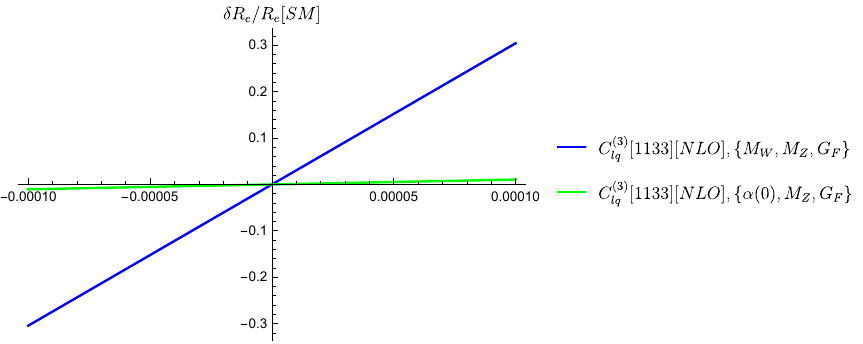}
	\caption{Results for $R_e$ proportional to $C_{lq}^{(3)}$ at LO and NLO in different input schemes normalized to the most accurate theoretical prediction. 
    } 
    \label{fg:Re3}
    \end{figure*}

\begin{figure*}
	\centering  
      \includegraphics[width=0.8\textwidth]{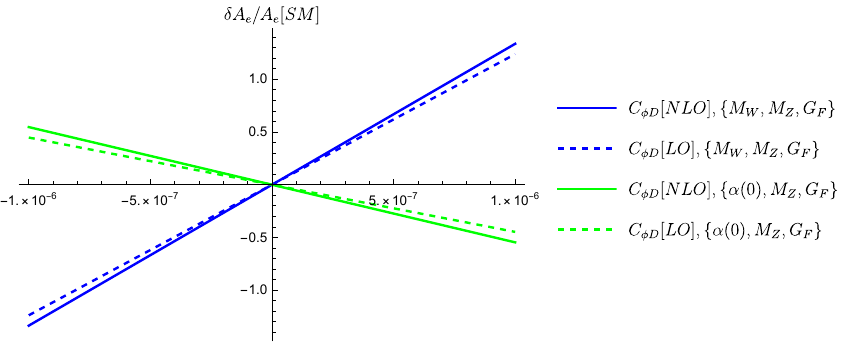}
	\caption{Results for $A_e$ proportional to $C_{\phi D}$ at LO and NLO in different input schemes normalized to the most accurate theoretical prediction. 
    } 
    \label{fg:Ae}
    \end{figure*}

\begin{figure*}
	\centering  
      \includegraphics[width=0.8\textwidth]{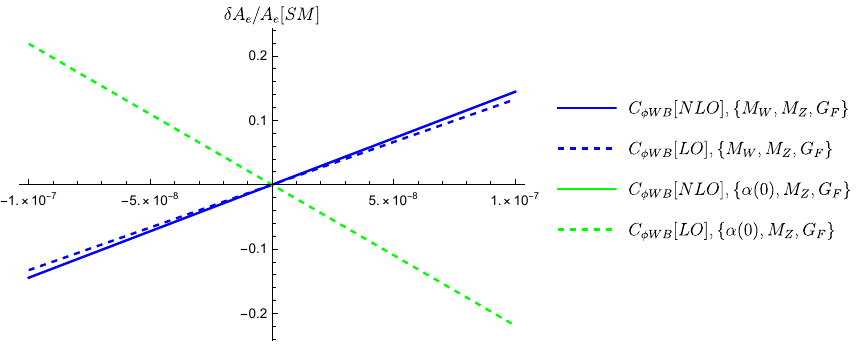}
	\caption{Results for $A_e$ proportional to  $C_{\phi WB}$ at LO and NLO in different input schemes normalized to the most accurate theoretical prediction. 
    } 
    \label{fg:Ae2}
    \end{figure*}

\begin{figure*}
	\centering  
      \includegraphics[width=0.8\textwidth]{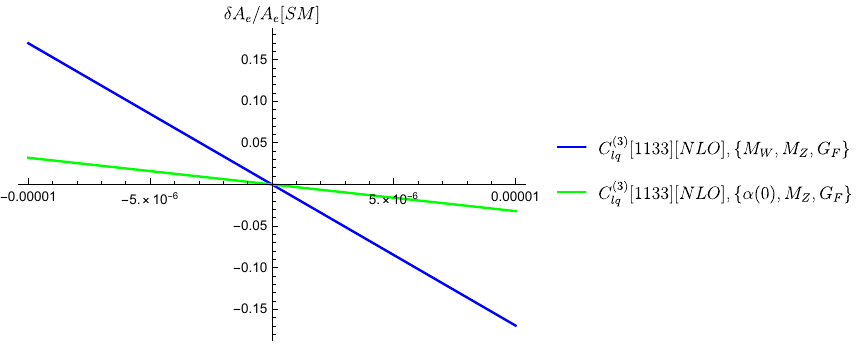}
	\caption{Results for $A_e$ proportional to  $C_{lq}^{(3)}$ at LO and NLO in different input schemes normalized to the most accurate theoretical prediction.  }
    \label{fg:Ae3}
    \end{figure*}

The input scheme dependence is known to be sizable for some operators.  We have upgraded the results of \cite{Bellafronte:2023amz} to include both the $\{\alpha(0), M_Z, G_F\}$ and $\{M_W, M_Z, G_F\}$ input schemes.  Other possible input schemes such as $\{\alpha(M_Z), M_Z, G_F\}$ or $\{G_F, sin\theta_{eff}^l, M_Z\}$  can be found in \cite{Biekotter:2025nln,Biekotter:2023vbh}. 
We show in Figs. \ref{fg:gamz}-\ref{fg:Ae3} results in different input schemes for some select coefficients where the difference between schemes is particularly large, using analogous notation to Eq. \eqref{eq:expa}.  The numerical values of the input parameters are given in the POPxf files.
Figs. \ref{fg:gamz} and \ref{fg:gamz2} show the contributions to the total $Z$ width that are proportional to $C_{\phi D}$ and $ C_{\phi WB }$ normalized to the most accurate theoretical prediction.  $C_{\phi D}$ and $C_{\phi WB}$ contribute at tree level and the NLO results do not differ significantly from the LO results. There is, however, a significant input scheme dependence.  The results for $C_{lq}^{(3)}$, which does not contribute at tree level, have a significantly smaller scheme dependence as seen in \ref{fg:gamz3}.  Figs. \ref{fg:Re}-\ref{fg:Re2} demonstrate that unlike the total width, the  dependence of $R_e$ shifts slightly going from LO to NLO for $C_{\phi D}$ and $ C_{\phi WB }$ and there is a significant scheme dependence that persists at NLO.  The scheme dependence of $R_e$ proportional to  $C_{lq}^{(3)}$ shown in Fig. \ref{fg:Re3} is smaller. Similar conclusions can be drawn from Figs. \ref{fg:Ae}-\ref{fg:Ae3}.  %The results can be used in global fits and to study the input parameter dependence of the results.

    
\subsection{Higgstrahlung}\label{sec:Higgstrahlung}

SMEFT results for the total cross sections for $e^+e^-\rightarrow ZH$ at center-of-mass energies $\sqrt{s}=240$, $365$ and $500$ GeV are also included as POPxf files using the calculations of \cite{Asteriadis:2024xuk,Asteriadis:2024xts}.  Both polarized and unpolarized results are presented. These files include the full LO and NLO SMEFT results including both QCD and electroweak corrections. The QED corrections can be found independently in \cite{Asteriadis:2024xuk,Asteriadis:2024xts} if needed.   The SMEFT contributions include all dimension-6 operators with no restrictions on the flavor structures and are accurate to linear order, ${\cal{O}}(\frac{1}{\Lambda^2})$, in the SMEFT expansion.  The arbitrary renormalization scale arising at NLO, $\mu$, is set to either $\mu$= 1 TeV or $\mu=\sqrt{s}$ and the input parameters are taken to be $M_W, M_Z$ and $G_F$. 

\section{Conclusion}
Global fits to SMEFT parameters provide an important source of information about potential high scale physics. Measurements of Higgs decays and Z-pole observables are crucial components of these fits.  
We have compiled results~\cite{GITLAB:nlo_smeft_popxf} that are accurate to linear order in the dimension-6 SMEFT and to NLO QCD and electroweak order for all 2- and 4- body Higgs decays and for all 2-body Z-pole decays using the POPxf data structure.  
These results can be straightforwardly implemented into global fits and experimental codes.  For completeness, we have also presented NLO SMEFT results for $e^+e^-\rightarrow ZH$ in the POPxf format.  

The eventual goal of the dimension-6 NLO QCD/EW SMEFT program is to have all the ingredients required to perform global fits consistently at NLO. At the LHC, most SMEFT predictions for production processes are currently available at NLO QCD accuracy, but not yet at full NLO EW precision. The results presented here therefore constitute an essential ingredient for future LHC fits at complete dimension-6 NLO QCD/EW accuracy. At the future FCC-ee, the Higgs decay results of this work can already be combined with calculations of the Higgstrahlung process and precision $Z$-pole observables to perform fits consistently at NLO QCD/EW order. In this setting, the POPxf formalism can play an important role by providing a common interface for the implementation and exchange of precision SMEFT predictions, making it easier to combine results from different calculations within global analyses.

\section*{Acknowledgements}

S.D. and C.D.P. are supported
by the U.S. Department of Energy under Contract No. DE-
SC0012704. P.P.G. is supported by the Ramón y Cajal grant~RYC2022-038517-I funded by MCIN/AEI/10.13039/501100011033 and by FSE+, and by the Spanish Research Agency (Agencia Estatal de Investigación) through the grant IFT Centro de Excelencia Severo Ochoa~No~CEX2020-001007-S. The work of L.B. is supported in part by the U.S.
Department of Energy under Grant No. DE-SC0010102 and by the College of Arts and Sciences of Florida State University. L.B. thanks the Technische Universität München (TUM) for the hospitality and the Excellence Cluster ORIGINS which is funded
by the Deutsche Forschungsgemeinschaft (DFG, German Research Foundation)
under Germany´s Excellence Strategy – EXC-2094 – 390783311 for
partial support during the completion of this work. Digital data in the POPxf format is provided at a \href{https://gitlab.com/mforslund/smeft-nlo-popxf}{GitLab~\faGitlab}~repository.

\bibliography{references.bib}

\end{document}